\documentclass[preprint]{aastex}

\newcommand{\laeq}{\raisebox{-0.5ex}{\,$\stackrel{<}{\sim}$}}

\shorttitle{Sub-arsecond structure of the outflow from DG TAU..}
\shortauthors{Bacciotti et al.}

\begin{document}

\title{HST/STIS spectroscopy of the optical outflow from DG Tau:
structure and kinematics on sub-arcsecond scales$^1$}


\author{Francesca Bacciotti\altaffilmark{2}, Reinhard Mundt\altaffilmark{3},
Thomas P. Ray\altaffilmark{2},\\ 
Jochen Eisl\"{o}ffel\altaffilmark{4},
Josef Solf\altaffilmark{4} and Max Camezind\altaffilmark{5}}

\altaffiltext{1}{Based on observations made with the 
NASA/ESA {\em Hubble Space Telescope}, obtained at 
the Space Telescope Science Institute, 
which is operated by the Association 
of Universities for Research in Astronomy,
Inc., under NASA contract NAS5-26555.} 
\altaffiltext{2}{School of Cosmic Physics, 
Dublin Institute for Advanced Studies,
5 Merrion Square, Dublin 2, Ireland.}
\altaffiltext{3}{Max-Planck-Institut f\"ur Astronomie, K\"onigstuhl 17,
D-69117  Heidelberg, Germany}
\altaffiltext{4}{Th\"uringer Landessternwarte Tautenburg,
Sternwarte 5, D-07778, Tautenburg, Germany}
\altaffiltext{5}{Landessternwarte K\"onigstuhl, D-69117 Heidelberg, Germany}

\begin{abstract}
We have carried out a spatio-kinematic study of the outflow from the 
classical T~Tauri star DG~Tau using the {\em Space Telescope 
Imaging Spectrograph} (STIS) on board the Hubble Space Telescope (HST). 
A series of seven spatially offset long-slit spectra 
spaced by 0.07$''$ were obtained along the 
axis of the outflow to build 
up a 3-D intensity-velocity ``cube'' in various forbidden emission lines 
(FELs) and H$\alpha$. Here we present high spatial resolution 
synthetic line images close to the star in distinct radial 
velocity intervals (from $\sim $  +50\,km\,s$^{-1}$ to $\sim$ -450\,km\,s$^{-1}$
in four bins, each $\sim$ 125\,km\,s$^{-1}$ wide). The lowest velocity emission
is also examined in finer detail (from +60\,km\,s$^{-1}$ to -70\,km\,s$^{-1}$ in
five bins $\sim$ 25\,km\,s$^{-1}$ wide).
We have found that the highest velocity and most highly collimated component, 
i.e.\ the jet, can be traced from DG~Tau to a distance D$\sim$0.7$''$. 
The jet is on the axis of a pear-shaped limb-brightened bubble which
extends between 0.4$''$ and 1.5$''$ from the source and which we interpret
as a bow shock. Other condensations are seen close to the star indicating 
ongoing temporal variations in the flow. The low-velocity component of the 
outflow is found to be spatially wide close to the source ($\sim$0.2$''$ at 
D=0.2$''$), in contrast to the high velocity jet (width $\laeq$0.1$''$). 
We have also found evidence to suggest that not only does the density in the 
outflow increase longitudinally with proximity to the source but that it 
also increases laterally towards the flow axis. Thus, at least in the case of 
DG~Tau, the flow becomes gradually denser as it increases in velocity and 
becomes more collimated. 
Our observations show a continous bracketing of the higher speed central
flow within the lower speed, less collimated, broader flow, down to 
the lowest velocity scales. This suggests that 
the low and high velocity FELs in the 
highly active T~Tauri star DG~Tau are intimately related. Implications 
of these observations for FEL models will be considered in a 
future paper \citep{brmesc00}.
\end{abstract}



\section{Introduction} \label{intro}

One of the most interesting questions in young stellar object (YSO) research 
is how their jets, e.g.\ \citet{camen97}, \citet{crete96} or \citet{eismundt97},
are collimated and accelerated. To address this problem one must obtain 
not only  high spatial resolution but in addition 
kinematic information as close as possible to their source. 
Ground-based long-slit spectroscopic studies of optically visible jet sources  
have shown that the structure and kinematics of the outflow region on scales 
$\laeq$1$''$ is rather complex (e.g.  \cite{solfbohm93}; 
\cite{hirthms97}). 
In a significant fraction of classical T~Tauri stars (CTTSs) the 
forbidden emission lines (FELs) show two (or more) blueshifted velocity 
components which have very different properties (note that in many CTTSs the 
corresponding redshifted part of the flow is occulted by a circumstellar 
disk, at least close to the source). In the case of double-peaked FEL profiles, 
the emission consists of a so-called high-velocity component (HVC), having 
typical radial velocities of -60 to -200 km\,s$^{\rm -1}$ and a 
low-velocity component (LVC), having typical radial velocities of -5 to 
-20 km\,s$^{\rm -1}$ with respect to the systemic velocity of the star (see 
\citet{hart95} and \citet{solf97} for comprehensive discussions). 
While the HVC can be spatially very extended (at least in the 
outflow direction) and is often identified with a jet, the LVC is much more 
compact ($\laeq$1$''$). Other differences between the two components
have also been noted, for example
the LVC appears to be of higher density but lower 
excitation than the HVC \citep{hirthms97} and the LVC shows a clear inverse 
correlation of velocity with increasing critical density. The latter 
effect has been interpreted as evidence of acceleration in the LVC 
with increasing distance from the star. 

A number of theories have been put forward to explain the origin of the 
various FEL components. For example Kwan and Tademaru (1988, 1995) 
have suggested that the LVC and HVC are separate flows. According to their 
model the LVC is a poorly collimated wind coming from the outer periphery 
of the YSO disk while the HVC is a separate jet launched from closer to the 
star. Others have sought to explain the observations in terms of a one
component, non-isotropic, wind model in which the appearence of separate 
FEL components is due to projection effects \citep{hartray89, saf93, ouypud94}. 
In any event, \citet{calvet97} has shown
that the luminosity of the two components is tightly correlated, which seems to
imply that they are not independent flows, and that the 
apparent dominance of one component over the other may be a density effect.    

In order to examine in more detail the nature of the compact FEL region,
 we have observed the CTTS DG~Tau with STIS on-board the HST. Multiple 
overlapping slit positions parallel to the outflow from this star were chosen 
so as to 
build up a 3-D spatial intensity-velocity ``cube''. Our target was picked not 
only because it is one of the closest CTTSs, but more significantly because 
its FEL region has a broad range of velocities,
probably also a result of the relatively small angle
between the jet axis and the line of sight
($\approx 38^{\circ}$, see \cite{eismundt98}).
This is important given the  moderate wavelength resolution of STIS.  
Historically, DG~Tau was amongst the first T~Tauri stars from which a jet-like 
outflow (HH~158) was discovered \citep{mundtfried83} and it has been imaged by 
the HST prior to the installation of the telescope's correcting optics 
\citep{kep93}. On large scales ($\approx$~10$''$) the jet seems to terminate 
in a bow shock \citep{eismundt98} while high resolution spectro-imaging 
ground-based studies \citep{lavalley97} have shown that the flow close to the 
star contains at least two resolved knots (at 2.7$''$ and 4$''$, epoch 1994.8). 
The outermost of these has a morphology and velocity gradient which is also 
consistent with it being a bow shock.

In \S 2 we describe our observational technique and give details of our STIS 
data. Our primary results are described in \S 3 and discussed in \S 4. A 
more detailed analysis will be presented in \citet{brmesc00}.  

\section{Observations} \label{obse}

STIS spectra of DG~Tau were taken on January 14 1999 using the G750M
grating and a central wavelength setting of 6581\,\AA . The spectral 
range of 562\,\AA\ included several strong forbidden lines
([OI]\,$\lambda\lambda$6300,6363, [NII]\,$\lambda\lambda$6548,6583, 
[SII]\,$\lambda\lambda$6716,6731) as well as H$\alpha$. 
The STIS/CCD detector, a 1024$\times$1024 pixel array, had 
a spectral resolution of 0.56\,\AA\,pixel$^{\rm -1}$ and a nominal 
sampling of 0.05$''$\,pixel$^{-1}$ in the dispersion and spatial 
directions respectively. The actual spatial resolution, 
however, was limited by the PSF of the HST in the red to a FWHM of 
approximately 0.1$''$.  The slit aperture was 52$\times$0.1 arcsec$^2$. 
Seven different long-slit spectra were taken, keeping the slit   
parallel to the outflow axis (P.A. 226$^{\circ}$), 
but with steps of 0.07$''$ in the transverse direction, i.e.\ with 
offsets southeast and northwest of the jet axis.
In this way we built-up a 3-D flux density/radial velocity 
data cube of the optical outflow from DG~Tau,
with a total spatial width of  about  0.5$''$  in the direction 
perpendicular to the jet axis.
The spectra are labelled S1,S2,...S7 going from the southeast to 
the north-west; the central slit position (S4) coincided with the star 
(RA~4$^{\rm h}$~27$^{\rm m}$~04$^{\rm s}$.71, Dec +26$^{\circ}$ 6$'$ 16.8$''$.) 
Exposure times were split in order to facilitate removal of 
cosmic rays.
Total exposure times were approximately 2740\,s for S1~--~S5, 1930\,s for 
S6 and 2050\,s for S7. Note that the somewhat shorter exposure times 
for the last two spectra were due to minor problems with the HST. 

The pipeline spectra were found to be contaminated by a large number of ``hot'' 
pixels and so the raw data were first fully re-calibrated using the {\em 
CALSTIS} suite of programs and new reference files made available by the STScI. 
Subsequent reduction and data analysis was carried out using standard IRAF
routines.  In order to minimise contrast problems close to the source, 
the stellar continuum, and that of a faint reflection nebula extending
up to 2$''$ from the source, were carefully 
subtracted from all images. This operation proved to be particularly critical 
for the central spectra S3, S4, and S5, due to the appearance of 
artificial undulations in the stellar continuum, caused by sampling 
effects (see the STIS Handbook v3.0).  

From the acquired spectra we created synthetic 2-D images 
for the various lines in four broad distinct radial velocity intervals
(Figs.\, 1--2). These images were constructed by adjoining the 
seven row-averaged columns of pixels obtained from S1--S7 for the 
corresponding velocity bin. Averaging was done over five columns, 
corresponding to approximately \ 2.83\AA , in the dispersion direction. 
In order to study the lowest velocity emission in finer detail,  
we also (Fig.\ 3) carried out a similar procedure in [OI]$\lambda$6300 and 
[SII]$\lambda$6731 but this time using just one column in the dispersion 
direction.  All velocities quoted in this {\em Letter} are systemic  
using v$_{\star, hel} \approx$ +16.5\,km\,s$^{\rm -1}$, 
as derived from the Li\,$\lambda$6707 photospheric absorption line.

\section{Results} \label{resu}

Our composite images of the outflow close to DG~Tau in the four broad velocity 
bins, and in each of four lines (H$\alpha$, [NII]$\lambda$6583, 
[SII]$\lambda$6731, and [OI]$\lambda$6363), are shown in Figs.\ 1 and 2. Note 
that we did not include [OI]$\lambda$6300 in these figures as 
the blueward portion of this line was missing from our spectra
due to a restricted choice for the STIS central wavelength. The velocity 
range is approximately +50\,km\,s$^{-1}$ to -450\,km\,s$^{-1}$ and the four 
velocity bins (low, medium, high, and very high), with widths of about 
125\,km\,s$^{-1}$, represent increasingly blueshifted velocities. On all 
figures the position of the star (i.e.\ the peak in the continuum subtracted 
light) is marked. Similar images, for the [OI]$\lambda$6300 and 
[SII]$\lambda$6731 lines, but only for the lowest velocities (from about 
+60\,km\,s$^{-1}$ to -70\,km\,s$^{-1}$) and with bin widths of only about 
25\,km\,s$^{-1}$, are shown in Fig.\ 3. 

First of all, we note that the appearence of the outflow  
changes, remarkably in some cases, from line to line  and from velocity bin
to velocity bin (see in  particular Figs.\ 1 and 2). 
The jet from DG~Tau is, as one might expect, most 
evident at the highest velocities and is visible as a narrow feature up to 
about 0.7$''$ from the star.  We will discuss it in more depth shortly. Aside 
from the jet a
limb-brightened bubble-like structure can also be observed that extends 
between 0.4$''$ and 1.5$''$ from DG~Tau. The ``bubble'', which is most 
evident at intermediate
velocities, in combination with the jet, gives an overall pear-shaped 
morphology to the emission. 

Comparison of Figs.\ 1, 2 and 3 lead to a number of important results:\\
(i) The lowest velocity emission close to the star ($\laeq$0.3$''$) 
is spatially broad, i.e.\ it is well resolved in the transverse direction 
to the outflow, but it does not extend far along the flow.  \\
(ii) Within $\laeq$0.5$''$ from the star, there appears to be a more or 
less gradual increase in the degree of collimation with velocity: at 
high and very high velocities, the flow is primarily confined to the central 
axis.\\
(iii) Beyond 0.5$''$ from the star there are at least two well-defined 
structures (which we have labelled A1 and A2 in the medium velocity 
H$\alpha$ image following the knot nomenclature of \citet{eismundt98}). The 
outermost one, A1, is bow-shaped
and observed in all lines but the innermost one, A2, is more easily seen in 
H$\alpha$. There is very high velocity gas immediately behind A1, which is seen 
in all lines but which is not symmetrically distributed 
around the jet axis.  The opposite effect is, incidentally, seen near A2 
where the higher velocity gas appears in the {\em downstream} direction. The 
emission from A2 is also distributed asymmetrically with respect to the outflow
axis.\\
(iv) Comparison of the [OI], [SII], and [NII] data (in Figs.\ 1 and 2) 
show that the central jet can be traced much closer to the source 
in [OI] and [NII] than in [SII]. 
This  result is probably  due to a combination 
of  quenching  and excitation effects (see \cite{brmesc00} for
details). Note that the [OI] line has a 100 (10) times higher 
critical density than the [SII] ([NII]) line, and that the 
[NII] line usually traces gas of higher excitation. 
The emission peak (centroid) in the [SII] and [OI] images 
moves outwards with increasing velocity. This effect is 
visible at all velocities, but it is most 
obvious at the highest velocities where virtually no  [SII] emission
and only faint [OI] emission is seen
in the region $\laeq$1$''$. Since  the high velocity jet
can be traced right back to the source in H$\alpha$, 
the lack of high-velocity and the presence of
low-velocity FEL emission at $<$ 1$''$ can probably only
be explained by quenching effects in {\em both} 
the longitudinal and lateral directions.
In other words our data suggest that the jet 
density increases not only with proximity 
to the star (in the longitudinal direction) but 
also transversely toward the outflow axis.
Without the latter effect one would not be able 
to explain why there is so much spatially extended and 
easily quenchable {\em low-velocity} FEL 
emission seen so close to the source.\\
(v) Quenching effects likewise manifest themselves at low velocities 
(see Fig.\ 3) in that the centroid of the low velocity [OI]$\lambda$6300
emission is closer to the star than the centroid of the [SII]$\lambda$6731
emission. This was expected on the basis of groundbased measurements
\citep{hirthms97}. \\
(vi) In the region close to the star ($\laeq$0.7$''$) [NII] emission 
comes primarily from the high velocity jet implying this is the region with 
the highest excitation. \\
(vii) The medium velocity bow-shaped structure A1 is less evident in [SII] 
and [OI] than in H$\alpha$, 
suggesting that its excitation level may be high. \\
(viii) There is a strong low velocity peak in H$\alpha$ emission 
coinciding with the star. This is almost certainly scattered emission from 
a region close to DG~Tau, including possibly a magnetospheric contribution 
\citep{edwards97} and therefore does not constitute part of the outflow 
{\em per se}. 

\section{Discussion} \label{disc}

The STIS observations presented here clearly show that the HVC emission in 
DG~Tau comes from the most highly focused part of the outflow i.e.\ the 
jet. The jet at high velocities can be traced back to at least 0.1$''$ 
(15\,AU) from the star where quenching effects become important 
even in the case of the 
[OI]$\lambda\lambda$6300,6363 lines, which have the highest critical density
($ \sim 10^6$ cm$^{-3}$) among the studied FELs. 
Obviously it is likely that the jet is collimated on even smaller scales. 
The shifts of the emission 
centroids in the different lines and in the different velocity bins can be 
explained quite naturally if there is an increase in jet density not only with 
proximity to the star (in the longitudinal direction) but also to the central 
outflow axis. That is to say the high velocity ``core'' of the jet is denser than its periphery. This is confirmed by an inspection of the 
[SII]$\lambda\lambda$6716,6731 doublet ratio (see Fig.\ 2 and Bacciotti et al.\
 2000).
Note also that structures such as A1 and A2 are reminiscent of the ``bubbles'' 
recently seen by \citet{krist99} in the case of XZ~Tauri. These are 
almost certainly internal working surfaces caused by 
temporal variations in the outflow from DG~Tau. Certainly there is 
plenty of evidence for strong jet velocity variations in 
DG~Tau (and in many other CTTS stars) on timescales of years. For example, 
the data of \citet{mundtetal87} and \citet{solfbohm93} show an increase in 
the radial velocity of the jet at D $\sim$ 0.5$''$  by a factor of 2 within 
8 years. Also the proper motion data of \citet{eismundt98} indicate large 
velocity variations. Finally, several of the knots in the DG~Tau jet show 
bow shock-like structures (see also the HST Archive data presented in 
\citet{brmesc00}) and these provide indirect evidence for strong 
velocity variations.\\ 
Turning now to the LVC, its nature still remains somewhat enigmatic. It was 
already clear from groundbased observations that the LVC and HVC differ in 
many properties such as density and excitation. A new difference is reported 
here i.e.\ the rather large 
spatial width W of the LVC perpendicular to the jet at distances D of about 
0.1$''$--0.3$''$ from the source. A comparison of W (at FWHM) between the 
high and low-velocity emission in [SII] and [OI] shows that the average W(LVC) 
$\sim$ 0.18$''$ while the HVC is hardly resolved at a distance D of 0.2$''$.
We note that this comparison of spatial widths can only be done using 
our [SII] and [OI] ``images'' as the H$\alpha$ data is heavily contaminated by 
stellar H$\alpha$ emission and in [NII]
the LVC is very weak. Another interesting result of our study
is the smaller velocity of the LVC at the edges of the flow (for full details 
see \citet{brmesc00}). For example at D\,=\,0.2$''$ 
the LVC peaks at $\sim$ -95\,km\,s$^{-1}$ in [OI] for the central slit 
position (S4) while in the two outermost slit positions (S1,S7) it peaks at  
about -18\,km\,s$^{-1}$. The corresponding values for the [SII] line are 
-60\,km\,s$^{-1}$ and -40\,km\,s$^{-1}$, respectively. Since these lines are
optically thin, such observations clearly point to a rise in the average LVC 
velocity as the central outflow axis is approached.\\ 
To what degree the observations of DG~Tau presented here, particularly of the 
LVC, are representative of other CTTSs is an open question. DG~Tau is one 
of the most active CTTSs known and we caution that the LVC of DG~Tau is 
unusual in that it has the highest absolute velocity of all the CTTSs listed 
by \citet{hart95}, as well as one of the highest accretion rates. That 
said it shares the typical properties of other LVCs and, in particular, 
the ratio of its luminosity (L) to that of the DG~Tau HVC is in perfect 
agreement with the L(HVC) v.\ L(LVC) relationship noted by \citet{calvet97}. 
Thus DG~Tau may simply be displaying the higher activity tail of the 
distribution of outflow properties amongst CTTSs. In conclusion our STIS observations show, for the first time, a quasi-continuous
variation in the outflow velocity close to a YSO in the transverse direction
to the flow. 
Detailed studies are required, however, to test whether these observations can 
constrain models for the generation of the LVC and HVC. 





\acknowledgments
We thank the anonymous referee for his/her very helpful comments.
FB was supported by an European Space Agency contract at the 
Dublin Institute for Advanced Studies during the course of this work.

\figcaption[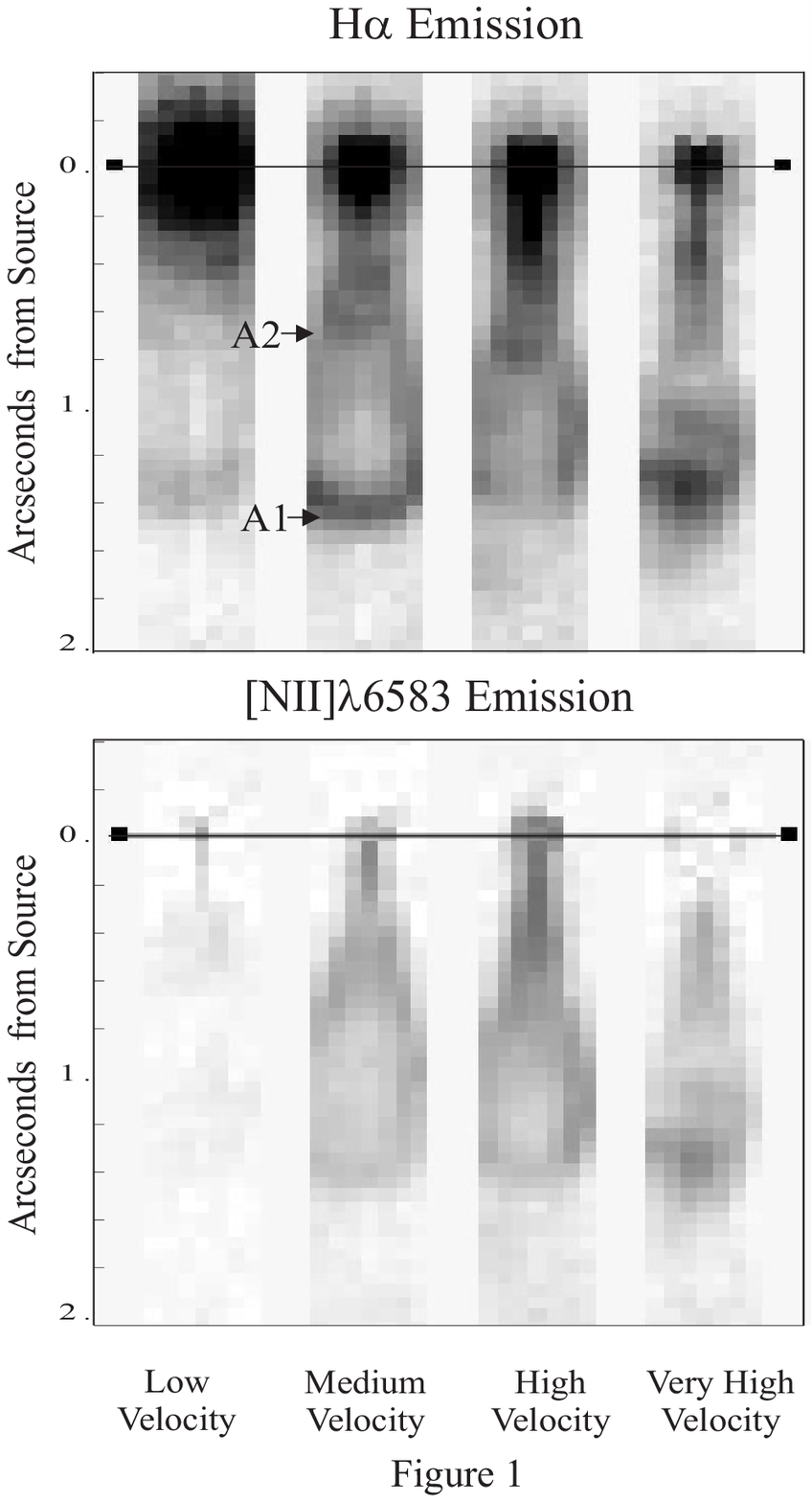]{Top panel: reconstructed H$\alpha$ 
images of the optical outflow close to DG~Tau in four distinct 
velocity bins: (from left to right) low from 
+59 to -67\,km\,s$^{-1}$, medium from -68 to 
-194\,km\,s$^{-1}$, high from -195 to -321\,km\,s$^{-1}$ 
and very high from -322 to -448\,km\,s$^{-1}$.
In this, and Figs.\,2 and 3, the flux density is 
displayed logarithmically from 10$^{-16}$ to 
10$^{-12}$\,erg\,s$^{-1}$\,arcsec$^{-2}$\,cm$^{-2}$\,\AA$^{-1}$.
The horizontal line marks the position of the star. 
The letters `A1' and `A2' mark two distinct structures in the flow, 
one of which, A1, is bow-like.
Bottom panel: same as top
panel, but for the [NII]$\lambda6583$ line. Velocity bins:
+55 -- -71 km\,s$^{-1}$, -72 -- -198 km\,s$^{-1}$,
-199 -- -325 km\,s$^{-1}$, and -326 -- -452 km\,s$^{-1}$.
\label{fig1}}

\figcaption[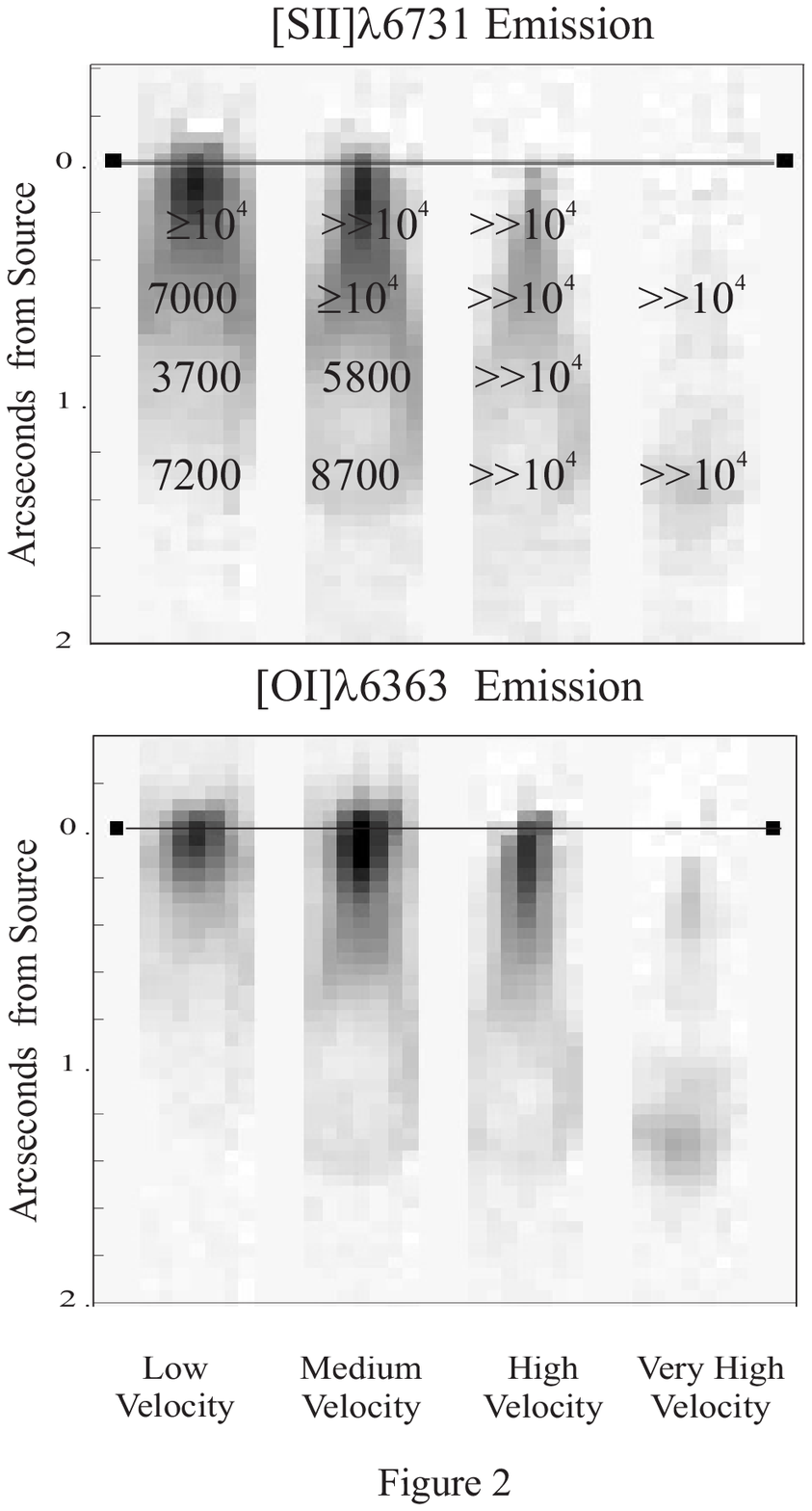]{Same as Fig.\,1, but in the light of
the [SII]$\lambda6731$ line (top panel) and [OI]$\lambda6363$ line
(bottom panel).  Velocity bins for the [SII] images:
+52 -- -71\,km\,s$^{-1}$,  -72 -- -195\,km\,s$^{-1}$,
-196 -- -319\,km\,s$^{-1}$, and -320 -- -443\,km\,s$^{-1}$.
The superposed numbers  refer to the corresponding average 
electron density values in cm$^{-3}$, calculated from the ratio of the [SII] 
lines. Velocity bins for the [OI] image:
+73 -- -57\,km\,s$^{-1}$, -58 -- -188\,km\,s$^{-1}$,
-189 -- -319\,km\,s$^{-1}$, and -320 -- -450\,km\,s$^{-1}$.
\label{fig2}}

\figcaption[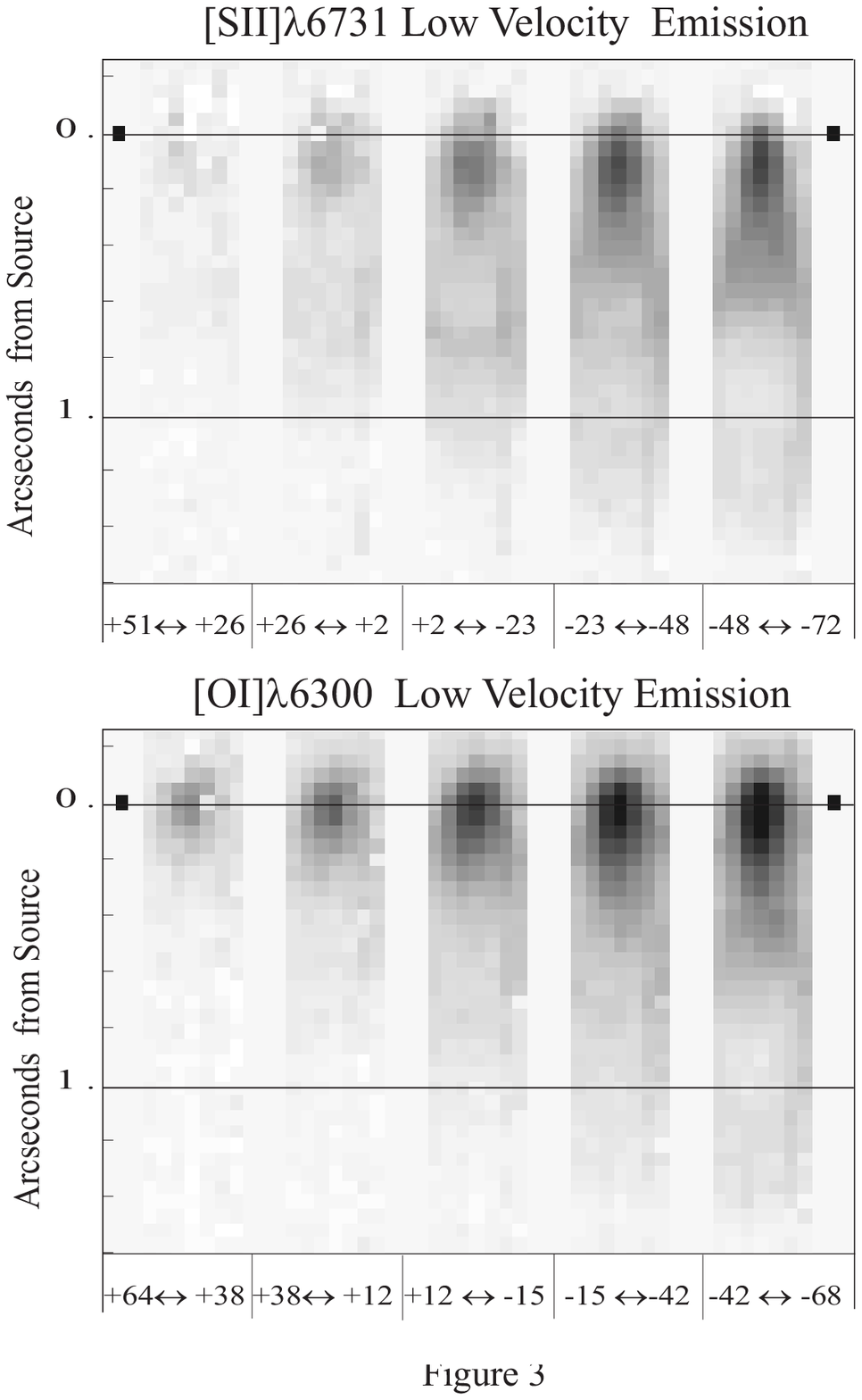]{Same as Fig.\,1, but 
only for the lowest velocities 
and in the light of the [SII]$\lambda6731$ line (top panel) and the
[OI]$\lambda6300$ line (bottom panel). Velocity bins for both images are given 
in km\,s$^{-1}$.
\label{fig3}}

\clearpage

\begin{figure}
\epsscale{1.13}
\vspace*{0.0cm}
\hspace*{-6.5cm}
\plotone{bacc.fig1.eps}
\end{figure}

\clearpage

\begin{figure}
\epsscale{2.7}
\vspace*{-4.cm}
\hspace*{-11.5cm}
\plotone{bacc.fig2.eps}
\end{figure}

\clearpage

\begin{figure}
\epsscale{0.8}
\vspace*{2.cm}
\hspace*{-1.3cm}
\plotone{bacc.fig3.eps}
\end{figure}

\end{document}